\documentclass[prl,twocolumn]{revtex4}
\usepackage{dcolumn}
\usepackage{multirow}
\usepackage{graphicx}
\usepackage{amssymb}
\usepackage{bm}
\usepackage{hyperref}
\usepackage{epstopdf}
\usepackage{color}
\usepackage{mathrsfs}
\usepackage{amsmath,amssymb,amsthm}
\usepackage{rotating}
\usepackage{sverb, longtable}
\usepackage{subfigure}

\usepackage[graphicx]{realboxes}
\usepackage{adjustbox}

\usepackage{footnote}
\begin{document}

\title{Did DESI DR2 truly reveal dynamical dark energy?}

\author{Deng Wang}
\email{dengwang@ific.uv.es}
\affiliation{Instituto de F\'{i}sica Corpuscular (CSIC-Universitat de Val\`{e}ncia), E-46980 Paterna, Spain}
\author{David Mota}
\affiliation{Institute of Theoretical Astrophysics, University of Oslo, P.O. Box 1029 Blindern, N-0315 Oslo, Norway}

\begin{abstract}
A fundamental question in cosmology is whether dark energy evolves over time, a topic that has gained prominence since the discovery of cosmic acceleration. Recently, the DESI collaboration has reported increasing evidence for evolving dark energy using combinations of cosmic microwave background (CMB), type Ia supernova (SN), and their new measurements of baryon acoustic oscillations (BAO). However, our analysis reveals that these combinations are problematic due to clear tensions among the CMB, BAO and SN datasets.
Consequently, DESI's claim of dynamical dark energy (DDE) is not robust.
A more reliable approach involves constraining the evolution of dark energy using each dataset independently. Through a statistical comparison for each dataset, on average, we find that DDE is strongly preferred over the $\Lambda$CDM model. This suggests that DDE likely exists, although its real parameter space remains elusive due to weak constraints on the dark energy equation of state and inconsistencies among the datasets.
Interestingly, when considering DDE, none of the individual datasets---including CMB, DESI DR2, Pantheon+, Union3, and DESY5---can independently detect cosmic acceleration at a significant level. Our findings not only clarify the current understanding of the nature of dark energy but also challenge the established discovery of cosmic acceleration and the long-held notion that dark energy exerts negative pressure. Both individual and combined datasets suggest that the ultimate fate of the universe is likely to be dominated by matter rather than dark energy.

\end{abstract}
\maketitle

\section{Introduction} 
The standard cosmological model, $\Lambda$ cold dark matter ($\Lambda$CDM), which has been confirmed by the CMB \cite{Planck:2018vyg,ACT:2025fju,SPT-3G:2022hvq,WMAP:2003elm,Planck:2013pxb}, BAO \cite{SDSS:2005xqv,2dFGRS:2005yhx,Beutler:2011hx,Blake:2011en,BOSS:2013rlg,BOSS:2016apd,eBOSS:2020yzd,deCarvalho:2017xye,eBOSS:2017cqx,eBOSS:2020gbb,DESI:2024uvr,DESI:2024lzq,DESI:2025zpo}, and SN observations \cite{SupernovaSearchTeam:1998fmf,SupernovaCosmologyProject:1998vns}, can successfully characterize various kinds of physical phenomena such as the cosmic acceleration on large scales and the clustering of matter on small scales \cite{Weinberg:2013agg}. However, it confronts two intractable challenges, namely the cosmological constant conundrum \cite{Weinberg:1988cp,Carroll:2000fy} and the coincidence problem, while suffering from the emergent cosmic tensions such as the so-called Hubble constant ($H_0$) tension and the matter fluctuation amplitude ($S_8$) discrepancy \cite{DiValentino:2020vhf,DiValentino:2020zio,DiValentino:2020vvd,Abdalla:2022yfr,DiValentino:2025sru}. It is very logically reasonable to query the validity of $\Lambda$CDM in verifying the fundamental theory and depicting the background dynamics and structure formation of the universe. So far, to solve these discrepancies, there are a great deal of alternative scenarios proposed by different authors (see \cite{Abdalla:2022yfr,DiValentino:2025sru} for reviews). It is worth noting that, besides the theoretical developments, more importantly, we require new independent and powerful probes with higher precision to give definite answers on some core puzzles. To achieve this goal, a very promising probe is BAO. 

BAO are regular and periodic matter density fluctuations of the universe \cite{Weinberg:2013agg}, which originate from sound waves induced by hot baryon-photon plasma before the epoch of recombination. The characteristic scale of BAO, approximately 150 Mpc, which is the maximum distance that the acoustic waves could travel in the primordial plasma before the plasma cooled to the point where it became neutral atoms at the recombination epoch, serves as a standard ruler in cosmology. Many BAO experiments such as 2dF \cite{2dFGRS:2005yhx}, 6dF \cite{Beutler:2011hx}, SDSS \cite{SDSS:2005xqv} and eBOSS \cite{eBOSS:2017cqx,eBOSS:2020gbb} map the late-time expansion history of the universe by measuring the apparent size of this ruler at different redshifts. BAO are very clean probes to explore the evolution of the universe over time, which is unaffected by the nonlinear physics on small scales and relatively robust to systematic uncertainties compared to other cosmological probes.

Recently, the DESI collaboration give the substantial evidence of DDE \cite{DESI:2024mwx}, based on their measurements of BAO in galaxy, quasar and Lyman-$\alpha$ forest tracers from the first data release (DR1) of the Dark Energy Spectroscopic Instrument (DESI) \cite{DESI:2024uvr,DESI:2024lzq}. Interestingly, this DDE evidence is enhanced \cite{DESI:2025zgx,DESI:2025fii} by the DESI's second data release (DR2) including more than 14 million galaxies and quasars, based on three years of operation \cite{DESI:2025zpo}.  

In theory, DDE predicts: (i) the equation of state (EoS) and energy densities of dark energy (DE) evolves over time; (ii) different expansion history of the universe from $\Lambda$CDM; (iii) the rate at which cosmic structures like galaxies and clusters form can be changed; (iv) the fate of the universe can be significantly affected. If DDE is finally demonstrated to be true, it will indicate that the vacuum is not empty and it does have matter. Therefore, the DESI's finding of DDE evidence is crucial for theory. Up to date, the addition of CMB and SN data to DESI DR2 leads to $2.8-4.2\,\sigma$ evidence of DDE \cite{DESI:2025zgx}, depending on which SN sample is used. Although the unprecedented precision and number of data points lead to the evidence of DDE, we should be very cautious about these results. The key reason is that the DDE evidence is derived by the data combination of CMB, DESI DR2 and SN, not from each probe independently. If each of CMB, DESI DR2 and SN gives consistent constraints on the cosmological parameters including the matter fraction and DE EoS, one can safely claim that DDE do exist in the late-time universe. Based on this concern, one has to seriously question: whether does DESI truly see DDE? Our results show that it is too early to claim the existence of DDE using the combinations of CMB, DESI DR2 and SN data, but independent datasets still give strong statistical preferences of DDE over $\Lambda$CDM. Future high precision observations can help clarify the status of DE.

\section{Basics} 
In the theory of general relativity \cite{Einstein:1916vd}, considering a homogeneous and isotropic universe, the Friedmann equations read as $H^2=(8\pi G\rho)/3$ and $\ddot{a}/a=-4\pi G(\rho+3p)/3$, where $a$ is the scale factor, $H$ is the cosmic expansion rate and $\rho$ and $p$ are the mean energy density and pressure of different species including radiation, baryons, dark matter and DE. Combining two Friedmann equations, one can express the dimensionless Hubble parameter $E(a)\equiv H(a)/H_0$ for a flat Chevallier-Polarski-Linder (CPL) universe \cite{Chevallier:2000qy,Linder:2002et} as
\begin{equation}
E(a)=\left[\Omega_{m}a^{-3}+(1-\Omega_{m})a^{-3(1+\omega_0+\omega_a)}\mathrm{e}^{3\omega_a(a-1)}\right]^{\frac{1}{2}}, \label{eq:ezcpl}
\end{equation}
where $\Omega_m$ is the matter fraction. It reduces to $\Lambda$CDM when $\omega_0=-1$ and $\omega_a=0$.   


\section{Data and methodology} 
We use the Planck 2018 high-$\ell$ \texttt{plik} temperature (TT) likelihood at multipoles  $30\leqslant\ell\leqslant2508$, polarization (EE) and their cross-correlation (TE) data at $30\leqslant\ell\leqslant1996$, and the low-$\ell$ TT \texttt{Commander} and \texttt{SimAll} EE likelihoods at $2\leqslant\ell\leqslant29$ \cite{Planck:2019nip}. We adopt conservatively the Planck lensing likelihood \cite{Planck:2018lbu} from \texttt{SMICA} maps at $8\leqslant\ell \leqslant400$. We use the most recent 13 BAO measurements from DESI DR2 including the BGS, LRG1, LRG2, LRG3+ELG1, ELG2, QSO and Ly$\alpha$ samples at effective redshifts $z_{\rm eff}=0.295$,  0.51, 0.706, 0.934, 1.321, 1.484 and $2.33$, respectively \cite{DESI:2025zgx,DESI:2025fii,DESI:2025zpo}. To completely explore the DE EoS at late times, we adopt three well-calibrated SN compilations: (i) Pantheon+ consisting of 1701 data points from 18 different surveys in $z\in[0.00122, 2.26137]$ \cite{Brout:2022vxf}; (ii) Union3 with 22 spline-interpolated data points derived by 2087 SN from 24 different surveys in $z\in[0.05, 2.26]$ \cite{Rubin:2023ovl}; (iii) DESY5 including 1735 effective data points in $z\in[0.025, 1.130]$ \cite{DES:2024jxu}. We carefully explain why we use calibrated SN datasets to implement constraints in the supplementary material (SM).

\begin{figure}[h!]
	\centering
	\includegraphics[scale=0.58]{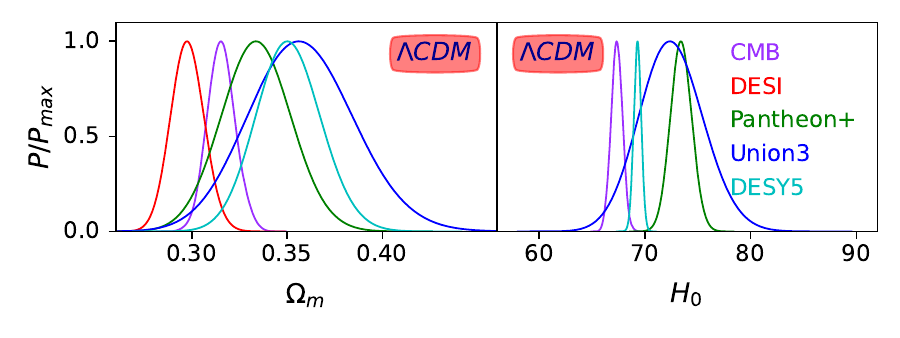}
	\includegraphics[scale=0.58]{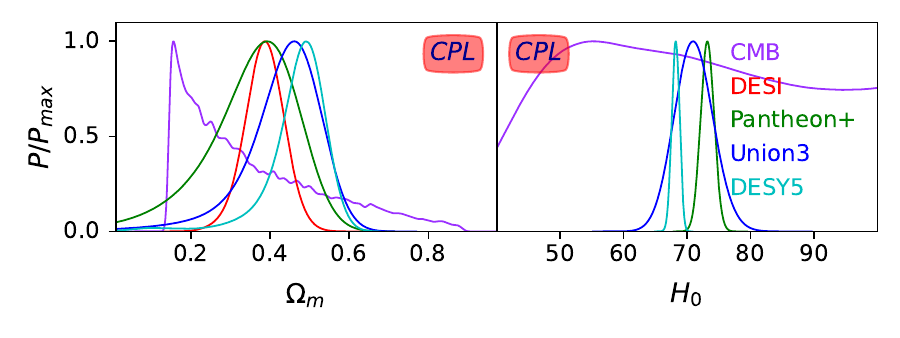}
	\caption{One-dimensional posterior distributions of the parameters $\Omega_m$ and $H_0$ from CMB, DESI DR2 and SN observations in the $\Lambda$CDM ({\it upper}) and CPL ({\it lower}) models, respectively. }\label{f1}
\end{figure}

\begin{figure*}
	\centering
	\includegraphics[scale=0.5]{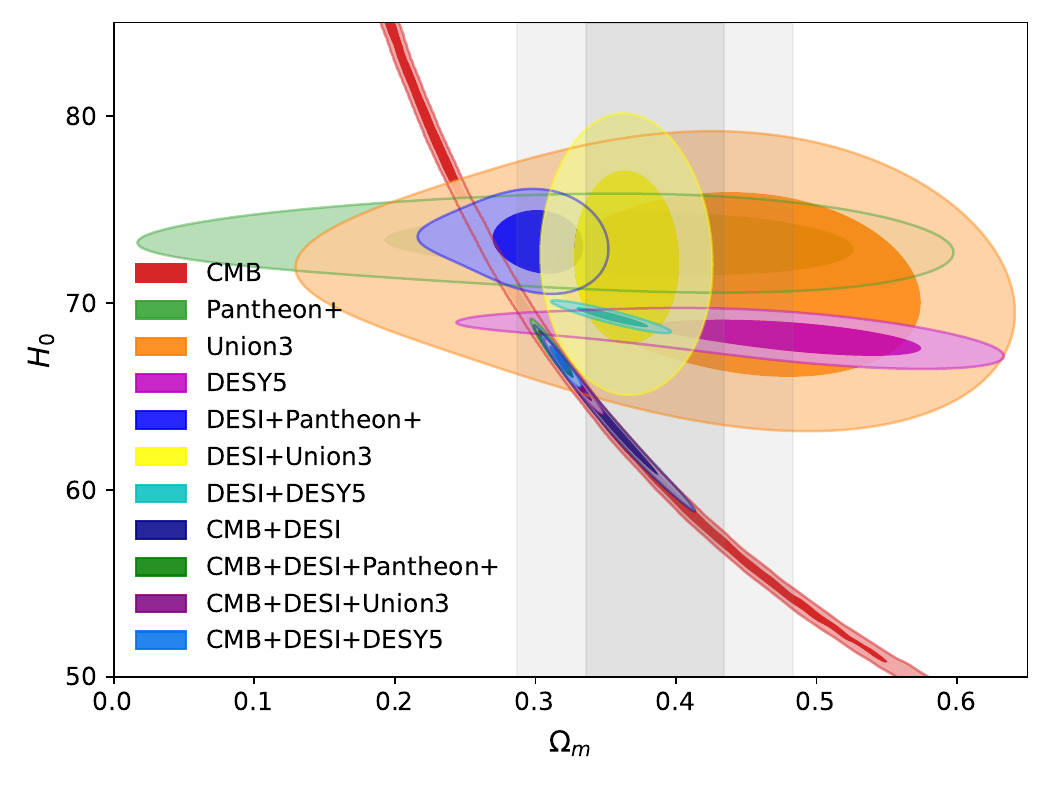}
	\includegraphics[scale=0.5]{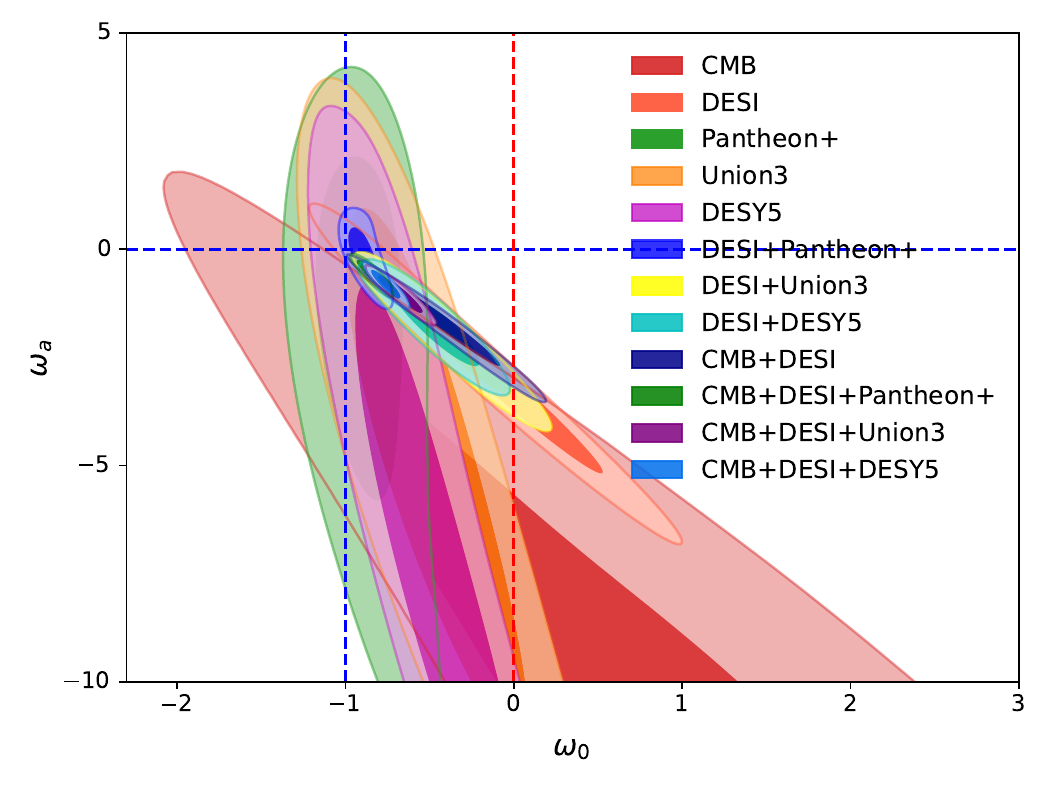}
	\caption{Two-dimensional posterior distributions of the parameter pairs ($\Omega_m$, $H_0$) and ($\omega_0$, $\omega_a$) from different datasets in the CPL model. The vertical shaded grey regions are the constrained $1\,\sigma$ and $2\,\sigma$ $\Omega_m$ values from DESI DR2. The cross point of blue dashed lines represents $\Lambda$CDM. The red dashed line denotes $\omega_0=0$.}\label{f2}
\end{figure*}

To calculate the background dynamics of the universe and theoretical power spectra, we use the Boltzmann solver \texttt{CAMB} \cite{Lewis:1999bs}. To implement the Bayesian analysis, we employ the Monte Carlo Markov Chain (MCMC) method to infer the posterior distributions of model parameters using the publicly available package \texttt{Cobaya} \cite{Torrado:2020dgo}. We assess the convergence of MCMC chains via the Gelman-Rubin criterion $R-1\lesssim 0.01$ \cite{Gelman:1992zz} and analyze them using \texttt{Getdist} \cite{Lewis:2019xzd}.

We use the following uniform priors for free parameters: the baryon fraction $\Omega_bh^2 \in [0.005, 0.1]$, cold dark matter fraction $\Omega_ch^2 \in [0.001, 0.99]$, acoustic angular scale at the recombination epoch $100\theta_{MC} \in [0.5, 10]$, scalar spectral index $n_s \in [0.8, 1.2]$, amplitude of the primordial scalar power spectrum $\ln(10^{10}A_s) \in [2, 4]$, optical depth $\tau \in [0.01, 0.8]$, present-day DE EoS $\omega_0 \in [-15, 20]$ and the amplitude of DE evolution $\omega_a \in [-30, 10]$. To produce a matter-dominated era at high redshifts, we impose the condition $\omega_0 + \omega_a < 0$ in the Bayesian analysis. The reason why we take such wide priors for the parameter pair ($\omega_0$, $\omega_a$) is that a large enough parameter space can completely present the constraining power of the DESI DR2 BAO measurements \cite{Wang:2024rjd}.


\section{$\Omega_m$ and $H_0$ tensions from independent probes} 
The DESI collaboration have noticed that there is an $\Omega_m$ tension among CMB, DESI DR2 and three SN samples under $\Lambda$CDM \cite{DESI:2025zgx}. SN datasets clearly prefer a larger matter fraction than CMB and DESI DR2. However, the accompanying $H_0$ tension is not reported. In Fig.\ref{f1}, we find that Pantheon+, Union3 and DESY5 exhibit $H_0$ tensions with CMB \cite{Planck:2018vyg} at $5.31\,\sigma$, $1.69\,\sigma$ and $3.03\,\sigma$ levels, respectively. Although three SN samples are well calibrated, they suffer from internal inconsistencies in their preferred $H_0$ values. For instance, Pantheon+ is in a $3.86\,\sigma$ tension with DESY5. The main goal of DESI is exploring the nature of DE including its possible dynamics. Based on the new motivation that DDE could help resolve the $\Omega_m$ tension in $\Lambda$CDM, we implement constraints on the CPL DDE and find that CMB prefers a smaller $\Omega_m$, while DESI DR2 and SN prefer larger matter fractions. Interestingly, $\Omega_m$ tensions between DESI DR2 and CMB, Pantheon+, Union3, and DESY5 are well alleviated from $1.57\,\sigma$, $1.83\,\sigma$, $2.10\,\sigma$ and $2.81\,\sigma$ to $0.14\,\sigma$, $0.21\,\sigma$, $0.49\,\sigma$ and $0.94\,\sigma$, respectively. It seems that DDE solves the $\Omega_m$ discrepancies, however, the price is DESI DR2 and SN give larger $\Omega_m$ (e.g. $\sim 0.5$) in CPL than in $\Lambda$CDM, while the $H_0$ tension between Pantheon+ and DESY5 increases from $3.86\,\sigma$ to $4.29\,\sigma$ (see Tab.\ref{tab:CPL}). Therefore, the well-known fact that the late universe is dominated by DE is largely challenged in CPL. Note that CMB cannot constrain $H_0$ in CPL, due to the weak constraint on the DE EoS.

\section{Inconsistencies from data combinations} 
The strong evidence for DDE primarily comes from the combination of CMB, BAO, and SN data. However, this result accommodates the inconsistencies among them. Hence, it is reasonable to question the validity of the combination. In Fig.\ref{f2}, the insensitivity of CMB to ($\omega_0,\,\omega_a$) leads to a much weaker constraint on $H_0$ and $\Omega_m$ in CPL than that in $\Lambda$CDM. Fortunately, the strong anti-correlation between $H_0$ and $\Omega_m$ remains. Combining CMB with BAO and SN, this anti-correlation will largely help compress the parameter space and consequently tighten constraints on ($\omega_0,\,\omega_a$). Unfortunately, there are beyond $1\,\sigma$ tensions between CMB and SN. Especially, DESY5 gives a $\sim 2\,\sigma$ tension with CMB. Even worse, the addition of DESI DR2 to SN leads to beyond $2\,\sigma$ tensions with CMB, particularly, DESY5 plus DESI DR2 gives a beyond $3\,\sigma$ tension with CMB. These tensions can also be captured in $r_d$ values. Adding DESI DR2 to Pantheon+, Union3 and DESY5 leads to a $5.30\,\sigma$, $2.86\,\sigma$ and $4.26\,\sigma$ tension with CMB when estimating $r_d$, respectively. Notice that there is a beyond $2\,\sigma$ tension between Pantheon+ plus DESI DR2 and DESY5 plus DESI DR2. This suggests a clear discrepancy among SN samples. Interestingly, CMB plus DESI DR2 is in beyond $3\,\sigma$, beyond $2\,\sigma$ and $\sim 1.5\,\sigma$ tensions with Pantheon+, DESY5 and Union3, respectively. All these discrepancies demonstrate that the DDE evidences from combined constraints are problematic. 

\begin{table*}[!t]
	\renewcommand\arraystretch{1.6}
	\begin{center}
		\caption{Mean values and $1\,\sigma$ (68\%) uncertainties of free parameters from different datasets in the CPL model. We quote the $2\,\sigma$ ($95\%$) upper limit of $\omega_a$ in the CMB-only case. The symbols ``$\bigstar$'' and $\blacklozenge$ denote unconstrained parameters by data and parameters with poor constraints, respectively. Here $h\equiv H_0/100$ in units of km s$^{-1}$ Mpc$^{-1}$. }
		\setlength{\tabcolsep}{2mm}{
			\label{tab:CPL}
			\begin{tabular}{l c c c c c c}
				\hline
				\hline
				Parameter & $\omega_0$ & $\omega_a$ & $\Omega_m$ & $H_0$ &$r_d$ & $hr_d$      \\
				\hline 
				CMB &  $2.4\pm 1.9$                 & $< -3.62$        & $0.344^{+0.054}_{-0.200}$ & $\blacklozenge$ &$147.32\pm 0.27$ & $103^{+20}_{-30}$  \\
				DESI &  $-0.17\pm 0.44$             & $-2.8\pm 1.6$    & $0.385\pm 0.049$         & $\bigstar$  & $\bigstar$  & $91.5^{+4.4}_{-4.9}$  \\
				Pantheon+ &  $-0.89\pm 0.17$        & $-2.1^{+3.2}_{-1.8}$ & $0.360^{+0.130}_{-0.086}$ & $73.2\pm 1.0$ &$\bigstar$ & $\bigstar$  \\
				Union3 &  $-0.45^{+0.28}_{-0.40}$     & $-5.4^{+4.7}_{-3.1}$ & $0.437^{+0.100}_{-0.066}$ & $71.1\pm 3.0$ &$\bigstar$  & $\bigstar$ \\
				DESY5 &  $-0.35^{+0.30}_{-0.41}$      & $-9.0^{+5.4}_{-4.5}$ & $0.471^{+0.075}_{-0.043}$ & $68.20\pm 0.60$ &$\bigstar$ & $\bigstar$  \\
				DESI+Pantheon+ &  $-0.885\pm 0.061$  & $-0.19\pm 0.46$ & $0.299^{+0.025}_{-0.016}$  & $73.3\pm 1.0$ &$136.1\pm 2.1$ & $99.72\pm 0.93$  \\
				DESI+Union3 &  $-0.37\pm 0.23$       & $-2.07\pm 0.82$ & $0.365\pm 0.024$ & $72.4\pm 3.0$ &$129.3\pm 6.3$ & $93.5\pm 2.6$ \\
				DESI+DESY5 &  $-0.47\pm 0.17$        & $-1.75\pm 0.61$ & $0.354\pm 0.016$ & $69.24\pm 0.34$ &$136.6\pm 2.5$ & $94.6\pm 2.1$  \\
				CMB+Pantheon+ &  $-0.868\pm 0.092$   & $-0.56^{+0.48}_{-0.42}$ & $0.311^{+0.011}_{-0.013}$ & $67.8\pm 1.2$ &$147.13\pm 0.25$ & $99.7\pm 1.8$  \\
				CMB+Union3 &  $-0.63\pm 0.14$        & $-1.39^{+0.68}_{-0.60}$ & $0.319^{+0.012}_{-0.014}$ & $66.9\pm 1.3$ &$147.17\pm 0.25$ & $98.5\pm 2.0$  \\
				CMB+DESY5 &  $-0.73\pm 0.10$          & $-1.04\pm 0.51$ & $0.3151^{+0.0094}_{-0.0110}$ & $67.3^{+1.10}_{-0.95}$ &$147.15\pm 0.25$ & $99.1^{+1.6}_{-1.4}$  \\
				CMB+DESI &  $-0.41^{+0.21}_{-0.25}$        & $-1.74^{+0.75}_{-0.59}$ & $0.353^{+0.022}_{-0.025}$    & $63.7\pm 2.0$ &$147.16\pm 0.23$ & $93.7\pm 3.1$  \\
				CMB+DESI+Pantheon+ &  $-0.843\pm 0.054$     & $-0.58^{+0.23}_{-0.19}$ & $0.3108\pm 0.0058$ & $67.62\pm 0.60$ &$147.28\pm 0.22$ & $99.59\pm 0.89$  \\
				CMB+DESI+Union3 &  $-0.673\pm 0.087$        & $-1.04^{+0.30}_{-0.27}$ & $0.3268\pm 0.0086$ & $66.02\pm 0.84$ &$147.22\pm 0.21$ & $97.2\pm 1.2$  \\
				CMB+DESI+DESY5 &  $-0.757\pm 0.057$         & $-0.82\pm 0.22$ & $0.3184\pm 0.0057$ & $66.86\pm 0.57$ &$147.24\pm 0.22$ &  $98.44\pm 0.84$  \\
				
				\hline
				\hline
		\end{tabular}}
	\end{center}
\end{table*}

Furthermore, we give more complete constraints on ($\omega_0,\,\omega_a$) using a wider prior than that the Planck and DESI collaborations used \cite{Planck:2018vyg,DESI:2025zgx,DESI:2025fii}. We find that CMB is in a $\sim 2\,\sigma$ tension with DESI DR2 and three SN samples are basically inconsistent with CMB at $\sim 1\,\sigma$ level. Since DESI DR2 with a clear different degeneracy direction of ($\omega_0,\,\omega_a$) from SN has a $\sim 1\,\sigma$ tension with SN, the DDE evidences from DESI DR2 plus CMB or SN are problematic. Additionally, even for the SN-only case, there is also an inconsistency among SN samples, i.e., Pantheon+ is consistent with $\Lambda$CDM within $1\,\sigma$ level, while Union3 and DESY5 exhibit beyond $1\,\sigma$ tensions with $\Lambda$CDM. Therefore, we should not place too much trust in the constraints derived from any pairwise combination of the datasets. Starting from this viewpoint, the reason why $\Omega_m$ values from the combination of CMB, DESI DR2 and SN are so close to $\Omega_m=0.3153\pm0.0073$ \cite{Planck:2018vyg} from the CMB-only constraint on $\Lambda$CDM should be a coincidence.  

\section{Solution} 
Due to the unprecedented precision and increasing number of data points, we should be cautious when studying possible new physics using the data combinations, which may bias the results away from the truth. Although there are discrepancies among CMB, BAO and SN data, an undeniable fact is that they all prefer the region of $\omega_0>-1$ and $\omega_a<0$. Especially, DESI DR2, giving a strong enough anti-correlation of ($\omega_0,\,\omega_a$), shows a $\sim 2\,\sigma$ preference with a relatively high precision. Concerning this, we implement a statistical comparison between CPL and $\Lambda$CDM. For CMB, we use the Bayesian evidence (see \cite{BE} for definitions) and find that the Bayesian factor $\ln B_{ij}=6.34 > 5$, indicating a strong evidence of CPL from CMB. For DESI DR2, Pantheon+, Union3 and DESY5, we employ the Bayesian information criterion (see \cite{BIC} for details) and obtain $\Delta \mathrm{BIC}=2.25$, 25.63, 3.92 and 11.49, respectively, indicating positive evidences from DESI DR2 and Union3 and strong evidences from Pantheon+ and DESY5 for DDE. Hence, DDE likely exists, however, we cannot determine the real parameter space of DDE according to such independent probes, due to their weak constraints on ($\omega_0,\,\omega_a$) and inconsistencies. If future observations (CMB, BAO or SN) with increasing precision can provide inconsistent constraints on DDE, we can safely claim its existence.       

\section{The fate of the universe} The existence of DDE will affect the composition, fate, expansion history and structure formation of the universe. The fact that three SN datasets allow $\Omega_m>0.5$ in CPL means that the late universe could be matter dominated not DE dominated. At least, CMB, DESI DR2 and SN all independently allow a large $\Omega_m\sim 0.5$. Interestingly, CMB, DESI DR2, Union3 and DESY5 do not rule out $\omega_0>0$ and $\omega_0>1/[3(\Omega_m-1)]$ \cite{CA}. Only Pantheon+ finds a beyond $1\,\sigma$ hint of late-time cosmic acceleration. One has to question whether the universe is accelerating now. CMB, DESI DR2, Union3 and DESY5 give, respectively, $\omega_0=2.4\pm 1.9$, $-0.17\pm 0.44$, $-0.45^{+0.28}_{-0.40}$ and $-0.35^{+0.30}_{-0.41}$, allowing that current universe could be slowing down or moving at a constant speed. Particularly, CMB gives a $1.3\,\sigma$ hint of $\omega_0>0$, allowing that DE has a positive pressure.  
Our findings not only profoundly challenge the understanding of cosmic acceleration, but also challenge the long-held notion that the pressure of DE is negative. Interestingly, we find that both independent and combined datasets prefer that the ultimate destiny of the universe is completely filled with matter not DE, i.e., $\Omega_m=1$. At some point in the distant future, matter will dominate the evolution of the universe. Finally, the universe will miraculously stop moving \cite{BT}. More details about the fate of the universe will be shown in \cite{Fate25}.

\section{Summary}
The DESI DR2 data, when combined with CMB and SN data, appears to provide increasing evidence for Dynamical Dark Energy (DDE). However, our analysis demonstrates that this conclusion is not robust. By constraining the CPL model with independent observations and various data combinations, we identify significant tensions between different datasets. These tensions lead to problematic constraints when incompatible datasets are combined.

Although the combined constraints yield $\Omega_m$ and $H_0$ values close to those from the CMB-only constraint on $\Lambda$CDM, we believe this to be coincidental. In the CPL model, the effect of the cosmological constant in $\Lambda$CDM is partially replaced by a larger matter fraction and DDE.

The combination of CMB, DESI DR2, and SN data results in a lower $H_0$ value because DESI DR2 indicates a high $\Omega_m = 0.385 \pm 0.049$. This high matter density directly leads to a low $H_0 = 63.7 \pm 2.0$ km s$^{-1}$ Mpc$^{-1}$ when combined with CMB data, due to the strong anti-correlation between $H_0$ and $\Omega_m$ in CMB observations. Given the crucial role of $H_0$ in the universe's background evolution, we explore the effects of $H_0$ and $H_0r_d$ on the parameters $\omega_0$ and $\omega_a$ for DESI DR2 in the SM. Additionally, we find that allowing for a free lensing amplitude $A_L$ can reduce the significance of DDE by approximately $1\sigma$, as $A_L$ and DDE are degenerate in the CMB lensing potential (see SM).

It is important to note that as the precision and volume of cosmological data increase, so do the tensions between different probes. This growing statistical complexity challenges our understanding of the data, underlying physics, and potential systematics. We are currently at a critical juncture in addressing these challenges.

\section{Acknowledgements} 
DW is supported by the CDEIGENT fellowship of Consejo Superior de Investigaciones Científicas (CSIC). DFM thanks the Research Council of Norway for their support and the resources provided by UNINETT Sigma2 -- the National Infrastructure for High-Performance Computing and Data Storage in Norway.

\clearpage

\appendix

\onecolumngrid
\section{\large Supplementary Material for ``Did DESI DR2 truly reveal dynamical dark energy?''}
\twocolumngrid

\onecolumngrid
{$\,\,$} In this supplementary material, first of all, we discuss the effect of calibration and marginalization of SN data on cosmological constraints. Then, we study the effect of CMB lensing amplitude $A_L$ on the DDE constraints using the combinations of CMB, DESI DR2 and SN observations. Since the DE EoS is closely related to the present-day cosmic expansion rate, we further investigate the effect of $H_0$ and $H_0r_d$ on the DDE constraints using only DESI DR2 BAO measurements. Finally, we compare the CMB-only constraints on the CPL DDE and $\Lambda$CDM.

\section*{A. On the calibration and marginalization of SN data}
Uncalibrated SN give an independent verification of cosmological parameters, helping to make sure that the results are not solely dependent on a specific calibration method or dataset. Using uncalibrated SN allows for consistency across different studies and datasets, making it easier to compare results and identify systematic uncertainties. If not considering the calibration, there will be a potential to discover new astrophysical phenomena or systematic effects that could affect dark energy measurements. However, there are a large number of disadvantages of using uncalibrated SN to constrain cosmological models. Uncalibrated SN can introduce systematic uncertainties due to variations in their intrinsic luminosity, which can influence the accuracy of cosmological measurements. There may be astrophysical biases, such as the dependence of SN luminosity on the properties of their host galaxies (e.g., mass, star formation rate), which can lead to inaccurate constraints on cosmological parameters. The precision of distance measurements may be limited due to the lack of calibration, leading to less reliable constraints on cosmological models. Without a proper calibration, there is a risk of overestimating or underestimating the effects of DE and the expansion rate of the universe. The analysis of uncalibrated SN often relies on assumptions about the properties of SN and their environments, which may not hold true across all redshifts. Uncalibrated SN do not account for intrinsic variations in SN properties, which can introduce additional noise into the datasets and reduce the accuracy of dark energy constraints.

Accordingly, for the calibration, there are systematic uncertainties associated with SN measurements. These can arise from variations in the SN properties, host galaxy environments, and dust extinction, which can affect the accuracy of distance measurements. The calibration of SN relies on other distance indicators, such as Cepheid variables \cite{Riess:2021jrx} or the Tip of the Red Giant Branch (TRGB) method \cite{Freedman:2021ahq}. Any uncertainties or biases in these calibrators can propagate into the SN distance measurements, affecting the estimation of $H_0$ and constraints on the DE parameters. The calibration process can be complex and time-consuming, requiring a careful consideration of various factors. The calibration process may depend on assumptions that do not hold true across all redshifts, affecting the accuracy of DE measurements.

\begin{figure}[h!]
	\centering
	\includegraphics[scale=0.55]{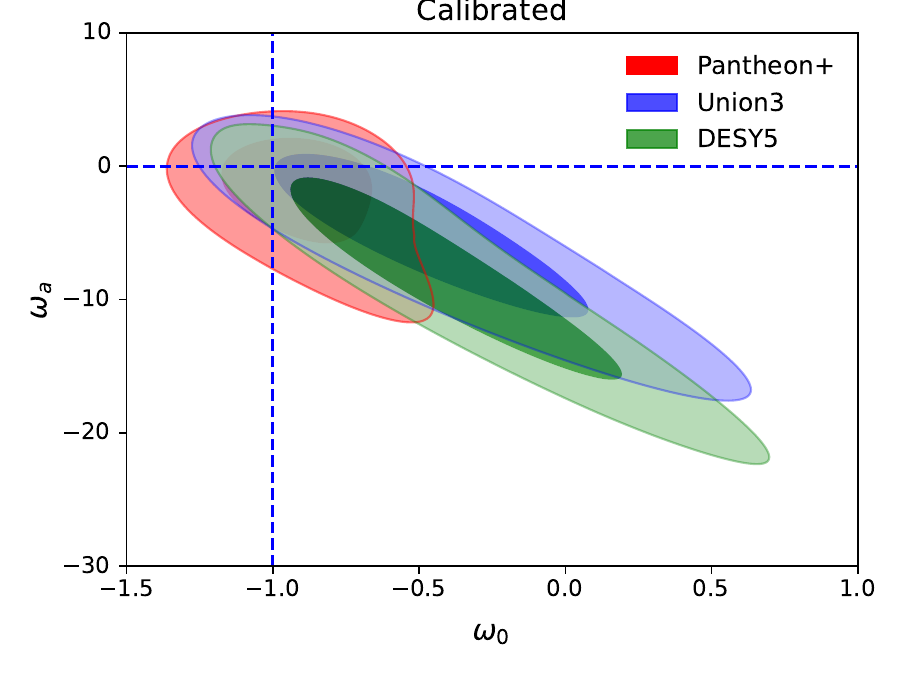}
	\includegraphics[scale=0.55]{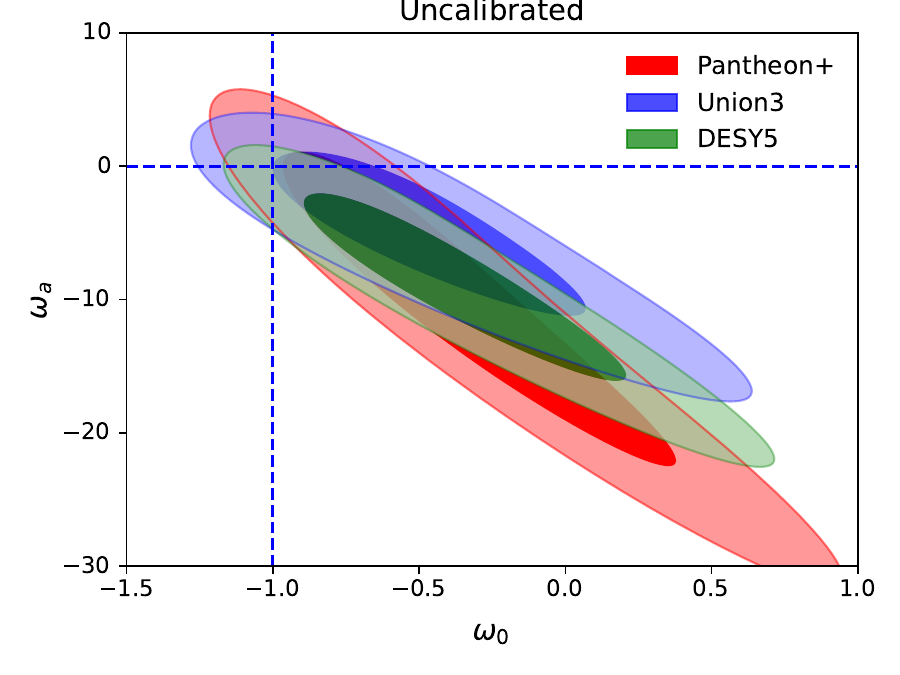}
	\caption{Two-dimensional posterior distributions of the parameter pairs ($\omega_0$, $\omega_a$) from calibrated and uncalibrated SN datsets in the CPL model. The cross point of blue dashed lines corresponds to $\Lambda$CDM.}\label{fs4}
\end{figure}

But, we emphasize that calibrated SN has the following advantages: (i) Calibrated SN provide more precise distance measurements, leading to more accurate constraints on DE models; (ii) Calibration can help to minimize systematic uncertainties, such as those arising from variations in SN luminosity and environmental factors; (iii) Calibrated SN reduce the dependence on assumptions about SN properties and their environments, leading to more robust constraints on DE models; 
(iv) Calibration accounts for environmental factors, such as dust extinction and host galaxy properties, reducing biases in DE measurements; (v) Calibrated SN account for intrinsic variations in SN properties, reducing noise in the datasets and enhancing the accuracy of DE constraints; (vi) Calibrated SN can be cross-validated with other independent datasets or methods, increasing the confidence in the findings; (vii) Calibrated SN are more sensitive to the effects of DE, allowing for more precise measurements of DE parameters; (viii) Calibration enables the detection of subtle changes in DE properties over time, enhancing the ability to study the evolution of DE.

In Fig.\ref{fs4}, one can easily find that the calibrated Pantheon+ SN sample gives a more conservative and tighter constraint on ($\omega_0$, $\omega_a$) than uncalibrated Pantheon+, while Union3 and DESY5 with and without calibration provide almost same constraints on the DE EoS. In this study, one of our main goals is digging up all the possible cosmological information embedded in each SN dataset. In light of this motivation, Pantheon+'s constraints on CPL, and the above advantages and disadvantages of uncalibrated and calibrated SN in constraining the DE models, we decide to use calibrated SN to provide more accurate and more conservative measurements on the DE parameters. Here ``conservative'' mainly means that the calibrated Pantheon+ data gives no deviation from $\Lambda$CDM.

It is worth noting that the released Union3 \cite{Rubin:2023ovl} and DESY5 \cite{DES:2024jxu} SN datasets in \texttt{Cobaya} \cite{Torrado:2020dgo} have been calibrated. Union3 and DESY5 give constraints on the cosmological parameters by marginalizing over the absolute magnitude $\mathcal{M}$. Actually, they provide $H_0=72.5\pm3.0$ and $69.31\pm0.35$ km s$^{-1}$ Mpc$^{-1}$ in the $\Lambda$CDM model, respectively. Even if taking a wide enough prior (say $H_0 \in [0,1000]$ km s$^{-1}$ Mpc$^{-1}$), their preferred $H_0$ values still do not change. For the consistency and completeness relative to Union3 and DESY5 samples, we also use the SH0ES calibration for Pantheon+ SN data \cite{Brout:2022vxf}. While marginalization can simplify models and focus on parameters of interest, it also has several potential disadvantages, especially when constraining a model with observations. Marginalizing over a parameter can lead to a loss of information about the relationships and dependencies between parameters. This can be problematic if the marginalized parameter is important for understanding the underlying processes in the model. By integrating out a parameter, the uncertainty associated with the remaining parameters could increase. This is because marginalization effectively spreads the probability mass over a wider range of values, reflecting the uncertainty about the marginalized parameter. The marginalized models can be harder to interpret because the impact of the marginalized parameter is no longer explicit. This can make it difficult to understand how changes in the marginalized parameter affect the overall model and its predictions. Since $H_0$ is closely correlated with ($\omega_0, \omega_a$) and $\Omega_m$, we think that an appropriate way should be not using the marginalization over $\mathcal{M}$ when implementing constraints. This suggests that SN distance observations and corresponding calibration data constructs a complete distance data system. Therefore, only when considering the SN calibration as a part of data, one can give a complete understanding of DDE. However, fortunately, here the marginalization of Union3 and DESY5 over $\mathcal{M}$ hardly changes their constraints on ($\omega_0, \omega_a$).


\begin{figure}[h!]
	\setcounter{figure}{0}
	\centering
	\includegraphics[scale=0.5]{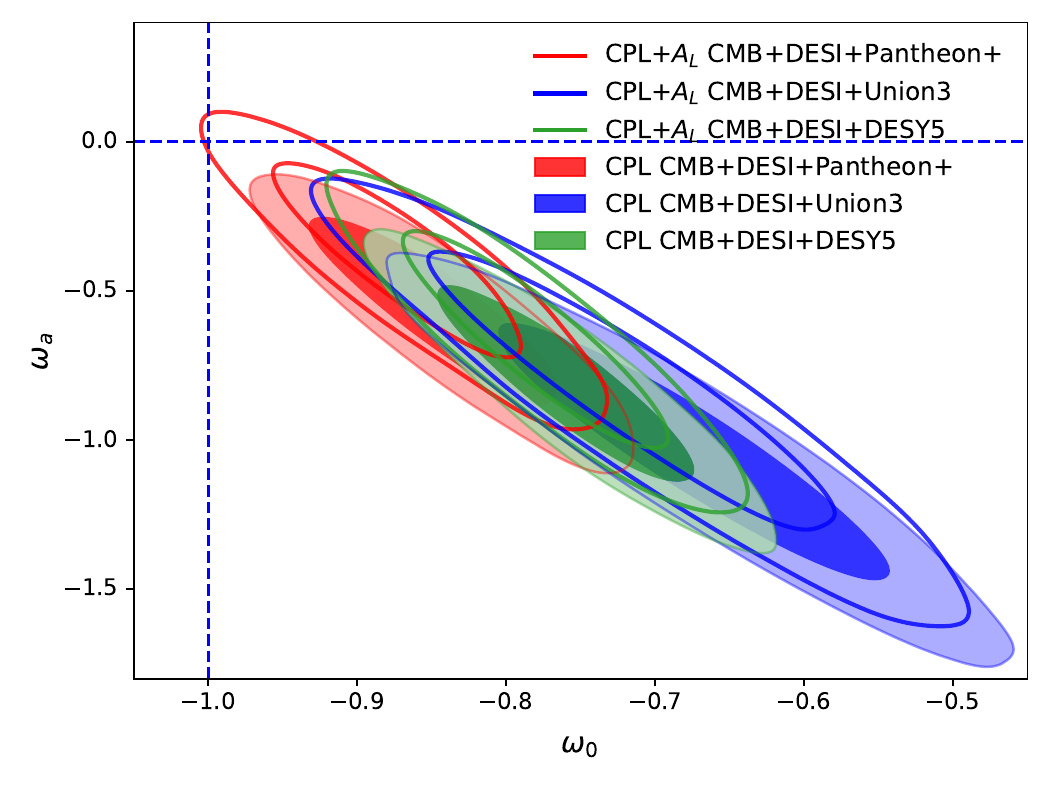}
	\caption{Two-dimensional posterior distributions of the parameter pairs ($\omega_0$, $\omega_a$) from CMB, DESI DR2 and SN observations in the CPL model with and without the lensing amplitude $A_L$. The cross point of blue dashed lines corresponds to $\Lambda$CDM.}\label{fs1}
\end{figure}

\begin{figure*}
	\centering
	\includegraphics[scale=0.43]{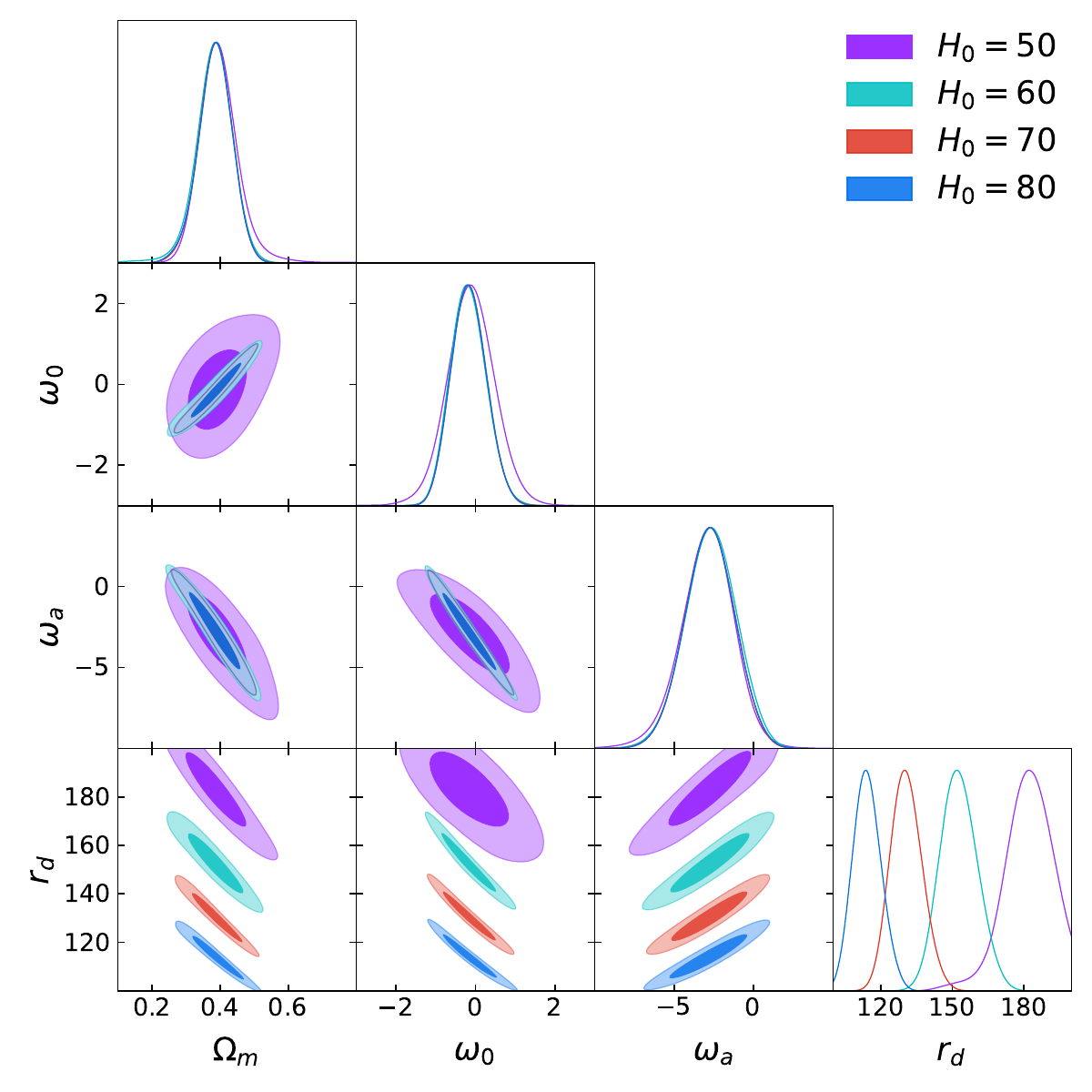}
	\includegraphics[scale=0.57]{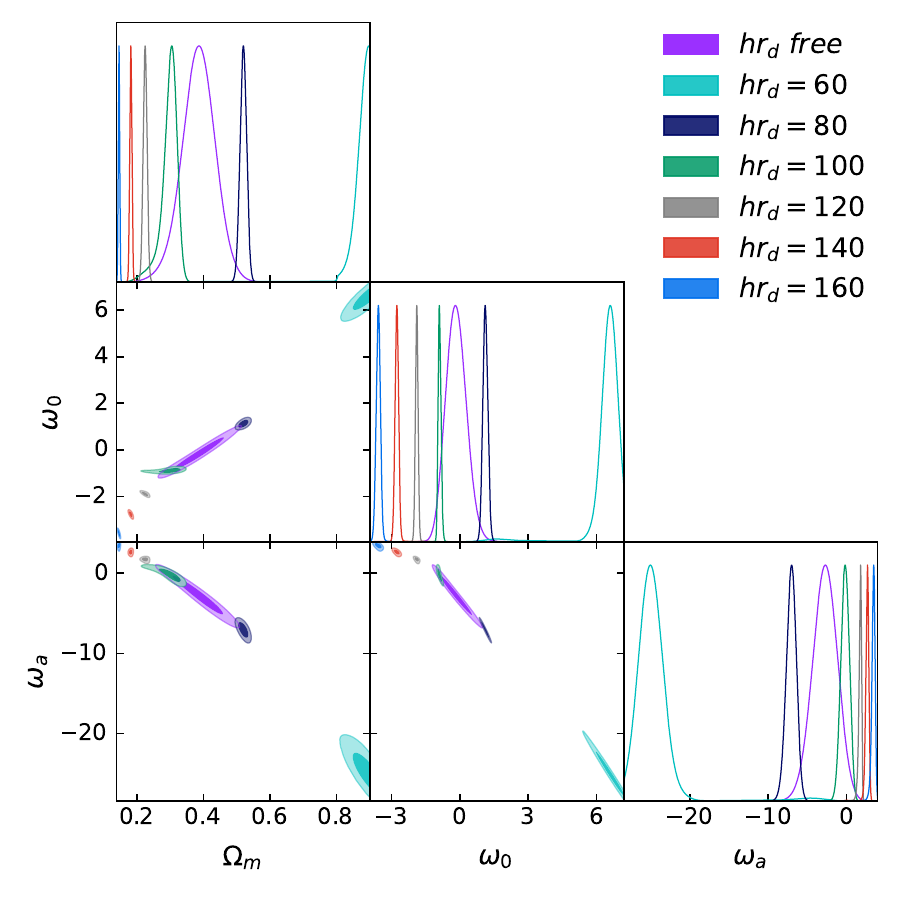}
	\caption{One-dimensional and two-dimensional posterior distributions of model parameters in the CPL model from the DESI DR2 data when considering different values of $H_0$ and $hr_d$, respectively. For the case of $hr_d$, we use a free $hr_d$ as a comparison.}\label{fs2}
\end{figure*}

\section*{B. The effect of lensing amplitude on DDE}
As we know, the CMB lensing anomaly is strongly degenerated with many new physics on cosmic scales, e.g., modified gravity \cite{Planck:2018vyg1,Wang:2023hyq}, cosmic curvature \cite{DiValentino:2019qzk,Handley:2019tkm} and effective field theory of DE \cite{Planck:2018vyg1}. Since both DDE and $A_L$ have impacts on the gravitational potential, considering $A_L$ could change the constraints on ($\omega_0$, $\omega_a$). Employing the data combination of CMB, DESI DR2 and SN, we find that a free $A_L$ can reduce a $\sim 1\,\sigma$ significance of DDE (see Fig.\ref{fs1}). This result holds for all three SN samples considered here.

\begin{figure*}
	\centering
	\includegraphics[scale=0.3]{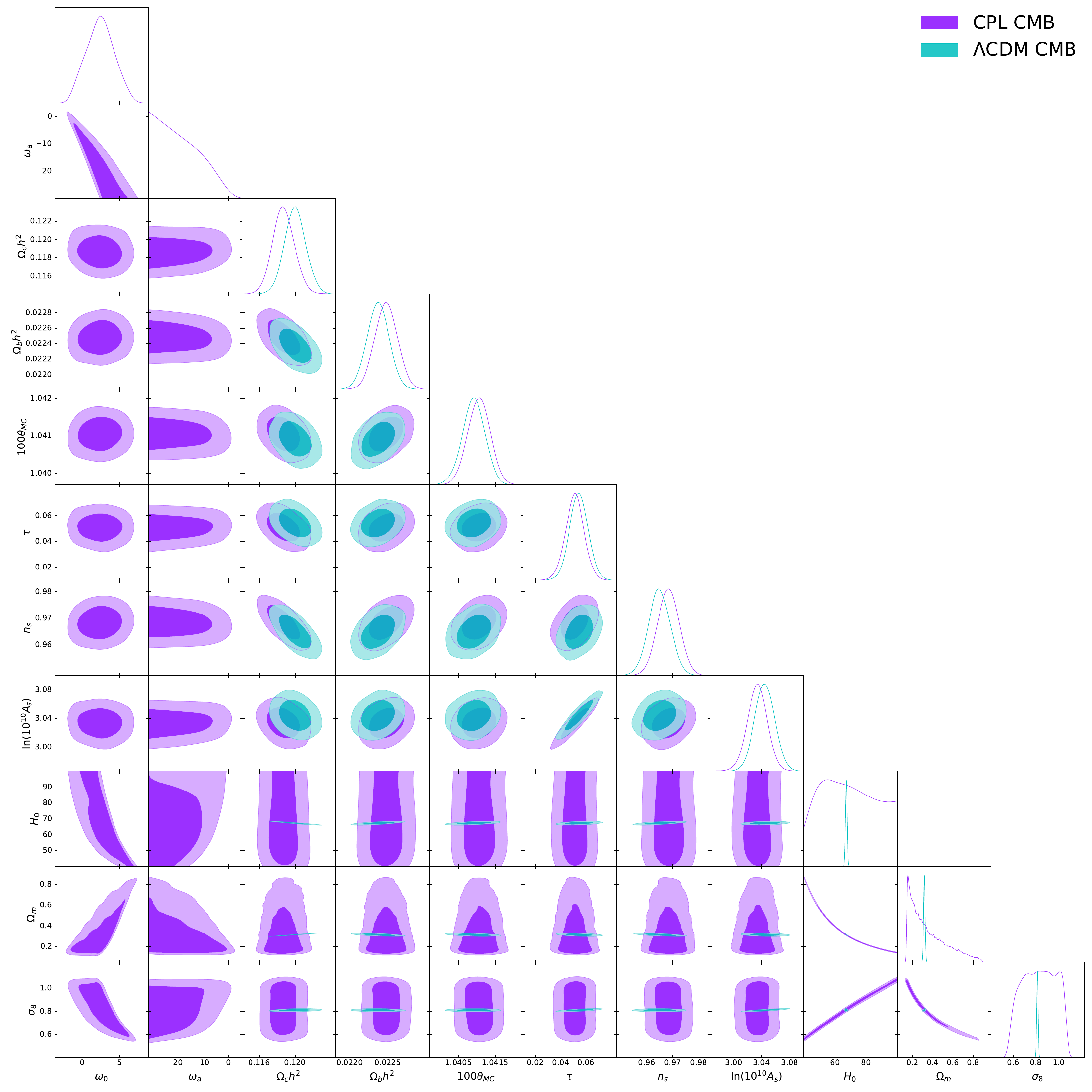}
	\caption{One-dimensional and two-dimensional posterior distributions of model parameters in the CPL and $\Lambda$CDM models from the CMB observations.}\label{fs3}
\end{figure*}

\section*{C. The effects of $H_0$ and $hr_d$ on DDE}
In theory, BAO require the information of the comoving sound horizon $r_d$ at the drag epoch from CMB to help determine $H_0$, since BAO cannot independently constrain $H_0$. However, it is interesting to see how the correlations between $r_d$ and the DE EoS (or matter density ratio) vary over different $H_0$ values. In Fig.\ref{fs2}, as expected, varying $H_0$ hardly changes the constraints on the matter fraction. Larger $H_0$ leads to smaller $r_d$. $r_d$ inherits the anti-correlations between $H_0$ and $\Omega_m$ (or $\omega_0$) and the positive correlation between $H_0$ and $\omega_a$.  It is worth noting that the whole parameter space will be enlarged when taking small $H_0$ values. Furthermore, we study how the compound parameter $H_0r_d$ impacts the DDE constraints. Similar to the case of $H_0$, larger $H_0r_d$ gives smaller $\Omega_m$ and $\omega_0$ as well as larger $\omega_a$. The constraints from the case of free $H_0r_d$ fall well in between those from $H_0r_d=80$ and 100, because it gives $H_0r_d=91.5^{+4.4}_{-4.9}$ (see the table in the main text).

\section*{D. CPL vs $\Lambda$CDM in light of CMB}
In Fig.\ref{fs3}, we make a comparison between the CPL DDE and $\Lambda$CDM models. Overall, DDE gives similar constraints on six basic parameters, although there are small shifts in some parameters. It is noteworthy that CMB is phenomenologically insensitive to the late-time DE EoS. This insensitivity leads to a poor constraint on ($\omega_0$, $\omega_a$). Therefore, one cannot well constrain the background quantities such as $H_0$ and $\Omega_m$ in the CPL DDE model.

\end{document}